\documentclass[traditabstract]{aa}
\usepackage[utf8]{inputenc}
\usepackage{amsmath}
\usepackage{natbib}
\usepackage{graphicx}
\usepackage{txfonts}
\usepackage{color}
\usepackage{multirow}
\usepackage{amssymb}
\usepackage{epsfig}
\usepackage{xspace}
\usepackage{microtype}
\bibpunct{(}{)}{;}{a}{}{,} 

\newcommand{\xper}{\object{X~Per}\xspace}
\newcommand{\foru}{\object{4U~2206$+$54}\xspace}
\newcommand{\axp}{\object{AXP~4U~0142$+$61}\xspace}
\newcommand{\xperl}{\object{4U~0352$+$309}\xspace}
\title{The hard X-ray emission of \xper.}
\author{V. Doroshenko\inst{1}, A. Santangelo\inst{1}, I. Kreykenbohm\inst{2,3}, R. Doroshenko\inst{1}}	
\institute{Institut für Astronomie und Astrophysik, Kepler Center for Astro and Particle Physics, Sand 1, 72076 Tübingen Germany\and
Dr. Karl Remeis-Sternwarte, Sternwartstrasse 7, 96049 Bamberg, Germany \and
Erlangen Centre for Astroparticle Physics (ECAP), Erwin-Rommel-Strasse 1, 91058 
Erlangen, Germany
}

\begin{document}

\bibliographystyle{aa}

\abstract{
We present an analysis of the spectral properties of the peculiar X-ray pulsar
\xper based on \rm{INTEGRAL} observations. We show that the source exhibits an
unusually hard spectrum and is confidently detected by \rm{ISGRI} up to more than 100\,keV. We find that two distinct components may be identified in the
broadband 4-200\,keV spectrum of the source. We interpret these components as the result of
thermal and bulk Comptonization in the vicinity of the neutron star and describe
them with several semi-phenomenological models. The previously reported absorption feature at
$\sim30$\,keV is not required in the proposed scenario and therefore its physical interpretation must be taken with caution. We also investigated the timing properties of the
source in the framework of existing torque theory, concluding that the observed phenomenology 
can be consistently explained if the magnetic field of the neutron star is $\sim10^{14}$\,G.}

\authorrunning{V. Doroshenko et al.}
\maketitle

\section{Introduction} 
\xperl is a persistent, low-luminosity, long periodic
accreting pulsar with a pulse period $\sim837$\,s and an X-ray luminosity
$L_{\rm x}\sim10^{35}{\rm\,erg\,s}^{-1}$ \citep{White:1976p4829}. The hard X-ray
spectrum and the observed variations of the spin-period
$\displaystyle|\dot{P}/{P}|\sim10^{-4}{\rm yr}^{-1}$
\citep{DelgadoMarti:2001p4574} imply
that the compact object is a neutron star.

The neutron star orbits the nearby \emph{Be} star \emph{X~Persei}
($d=0.95\pm0.2$\,kpc, \citealp{Telting:1998p4827}). 
The binary orbit is wide and almost circular, with an orbital period of about
250\,d and eccentricity $e\sim 0.11$ \citep{DelgadoMarti:2001p4574}. The
compact companion orbits relatively far away ($\sim2$\,AU,
\citealp{Levine:12p4830,DelgadoMarti:2001p4574}) from the optical companion
and does not pass through the disk of the \emph{Be} star. Consequently, the
source does not exhibit the outbursts at periastron typical for this class of
sources. The observed X-ray luminosity is, however, three orders of magnitude
higher \citep{DelgadoMarti:2001p4574} than what is expected for
accretion from a fast ($\sim800$\,km\,s$^{-1}$) low-density stellar wind
\citep{HammerschlagHensberge:1980p4826,Bernacca:1981p4824}. Indeed, to explain
the observed X-ray luminosity, \cite{DelgadoMarti:2001p4574} suggested that
accretion proceeds from a slower ($\sim150\,{\rm km\,s}^{-1}$) and denser wind
extending from the circumstellar disk of the \emph{Be} companion.

Another unusual aspect of \xperl is its hard X-ray spectrum. Most accreting
pulsars exhibit a power-law spectrum with a cutoff above $E \sim 20$\,keV. \xper spectra in the hard energy range
($\ge10$\,keV) have been typically fitted with a thin thermal bremmstrahlung model 
with $kT\sim10$\,keV, sometimes with an additional hard-energy tail
\citep{White:1976p4829,Mushotzky:1977p4834, Frontera:1979p4833,
Worrall:1981p4835}. The ``standard'' cutoff power-law model has also
been used \citep{Frontera:1979p4833,White:1983p4837} although with an unusually
low ($\sim1$\,keV) cutoff energy.

The broadband X-ray spectrum of \xper is more puzzling.
To describe the broadband (0.1-200\,keV) \rm{BeppoSax} spectrum, \cite{diSalvo:1998p4570} used a model consisting of two power-law components.
The first dominates at lower energies and is modified by an exponential cutoff at $\sim2$\,keV. The second is harder and is characterized by a low-energy turnover and a cutoff at
$\sim65$\,keV. \cite{diSalvo:1998p4570} argued that the partially thermalized emission from the atmosphere of the neutron star is
responsible for the soft part of the spectrum. The hard part was
interpreted either as cyclotron emission \citep{Nelson:1995p5112} in a strong
magnetic field, or was caused by high-temperature ($\sim90$\,keV) thermal
bremmstrahlung on electrons accelerated in a collisionless shock above the
surface of the neutron star. \cite{Coburn:2001p4571} described the broadband
(4-120\,keV) \rm{RXTE} spectrum of the source as a combination of a blackbody
at low energies ($kT\sim1.8$\,keV) and a power law modified by a broad
absorption feature at $\sim30$\,keV, which the authors interpreted
as a cyclotron resonance scattering feature (CRSF). The magnetic field was therefore estimated to be $B\sim2.6\times10^{12}$\,G.

The spin period of \xperl is peculiarly long. As expected for wind accretion,
it varies erratically, although both spin-up and spin-down trends with
$\displaystyle|\dot{P}/{P}|\sim10^{-4}{\rm yr}^{-1}$ have been identified
\citep{DelgadoMarti:2001p4574}. The pulse-profile is single-peaked and
sinusoidal. It changes very little with energy, although there is some
hardening at peak minimum and maximum
\citep{diSalvo:1998p4570,Coburn:2001p4571}.

In this paper we focus on the spectral properties of the source as observed by
\rm{INTEGRAL}. We show how the broadband continuum spectrum of \xper is
unusually hard and has two distinct humps, which we interpret as the result of
thermal and bulk Comptonization in the accretion flow close to the surface of
the neutron star. Similarities with other sources are also discussed. 

\section{Data and analysis} The International Gamma-Ray Astronomy Laboratory
(\rm{INTEGRAL}), launched in October~2002 by the European Space Agency
(ESA) is equipped with theree co-aligned coded mask instruments: \rm{ISGRI}
(Integral Soft Gamma Ray Imager,~\citealt{Ubertini:2003p1120}), \rm{JEM-X} (Joint European X-ray
Monitor,~\citealt{Lund:2003p1129}) and \rm{SPI} (Spectrometer on
\rm{INTEGRAL},~\citealt{Vedrenne:2003p1138}).

To measure the broadband spectrum of the source we used the available
archival data with \xper within the full-coded field of view of the \rm{JEM-X}
instrument (which has the smallest field of view among three instruments). 
This results in a total effective exposure of about $400\,$ks in 295 pointings from MJD~53948 to 54874. The data were reduced with the standard software
OSA-9.0 and the set of calibration files IC-9.0 provided by
ISDC\footnote{http://www.isdc.unige.ch/integral/analysis}. For \rm{SPI}
imaging we used the \emph{spiros} branch of the analysis pipeline.

The \rm{ISGRI} image immidiately revealed something peculiar: unlike
most other accreting pulsars, \xper was confidently detected (at 8.5$\sigma$) in the 100-200\,keV
energy range. Indeed, in this energy range \xper is the only source
detected with significance greater than $5\sigma$ in the \rm{ISGRI} field of view. 
This is not because the source is close to us. If other nearby
\emph{Be} X-ray binaries (with $d\sim2-3$\,kpc like Her~X-1 or Vela~X$-$1)
were as hard as \xperl, they would be just a factor of 10 weaker and therefore easily
detectable with the available \rm{ISGRI} exposures. 

\begin{figure}[t]
	\centering
		\includegraphics[width=0.45\textwidth]{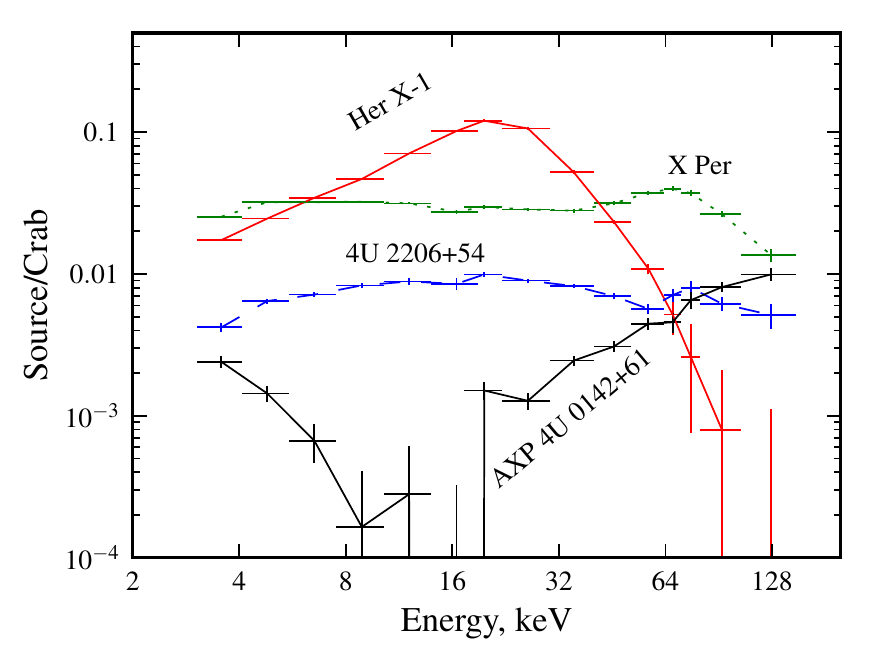}
	\caption{Crab-normalized broadband spectra for Her~X-1 (a typical accreting pulsar), \xper, \foru (a source that we argue is similar to X~Per), AXP~4U~0142+61 (brightest of the anomalous X-Ray pulsars). Note how hard the spectrum of X~Per is compared to Her~X-1.}
	\label{fig:versus}
\end{figure}

The comparison of the broadband spectrum of X Per with that of other pulsars
(see Fig.~\ref{fig:versus}) shows that X Per~is relatively bright at low
energies, is fainter than a typical pulsar in the 8-40 keV range, but
outshines its peers above $\sim50$\,keV.\footnote{for this comparison we used the
quick-look data provided by ISDC HEAVENS \citep{heavens}}
To perform a detailed spectral analysis we extracted the pulse-phase-averaged spectrum of
X~Per using data from the \rm{ISGRI}, \rm{SPI}, and \rm{JEM-X1}
instruments (\rm{JEM-X2} was switched off during most of the observations,
therefore there were few usable data).

\begin{figure*}[t]
	\centering
	\vspace{-10pt}
		\includegraphics[width=0.95\textwidth]{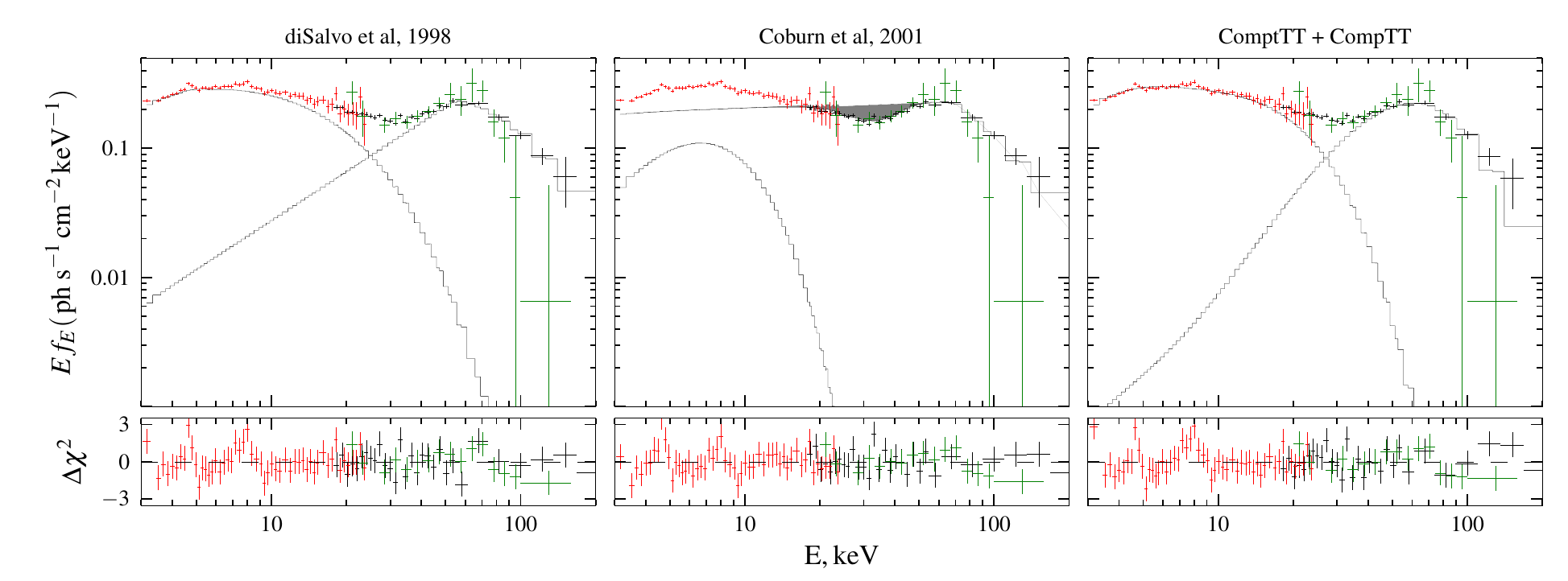}
	\caption{$Ef_E$ spectrum of \xper observed by INTEGRAL. The best-fit spectra of the three different models discussed in the text and their residuals are shown.
	Data from \rm{ISGRI} (black), \rm{JEM-X} (red) and \rm{SPI} (green) instruments were used. The contribution
	of the single model components is also shown. For the model by \cite{Coburn:2001p4571} (middle pane), we also show (shadowed area) the effect of
	including an absorption-like line, which is required by the specific continuum.}
	\label{fig:spe}
\end{figure*}

To describe the spectrum we first used slightly modified versions of the
models used by \cite{diSalvo:1998p4570}\footnote{no rollover for the hard component
is required} and by \cite{Coburn:2001p4571}\footnote{cutoff for the hard component
is required}. Both models aim to mimic characteristic features of a Comptonization spectrum, therefore we also attempted to use a true Comptonization model derived from first principles \citep{Titarchuk:1994p2324} with two components 
of independent temperatures and optical depths.  In all cases photoelectric
absorption was included with the column depth fixed to
$2\times10^{21}{\rm\,atoms\,cm}^{-2}$
\citep{Haberl:1994p5113,diSalvo:1998p4570} since \rm{INTEGRAL} does not have the
low-energy coverage required to constrain it. The results are summarized in
Table~\ref{tab:allspe} and Figure~\ref{fig:spe}.

The best-fit values of the corresponding models are consistent with those reported previously by \cite{Coburn:2001p4571} and
\cite{diSalvo:1998p4570}. Independently of the model used, two distinct
``humps'' are identified in the $Ef_E$ spectrum (see Fig.~\ref{fig:spe}). At
lower energies the spectrum is similar to that typically observed in other
HMXBs. It can be described as a cutoff power law, although the cutoff energy is
quite low. At higher energies, however, a second ``hump'' starts to dominate
and makes the spectrum unusually hard with a cutoff energy around 60\,keV.
\begin{table*}
	\begin{center}
	\begin{tabular}{l@{\hspace{0.3cm}}lllll@{\hspace{1.6cm}}llll}
\hline
\hline
   {\sl Model}          & \multicolumn{5}{c}{{\sl Low energy part}} & \multicolumn{4}{c}{{\sl High energy part}} \\ \hline\\[-2ex]

\multirow{2}{*}{\cite{diSalvo:1998p4570}} &  $E_{\rm cut, low}^\dagger$ & $E_{\rm fold, low}^\dagger$ &         $\Gamma_{\rm low}$ &           &         $A_{\rm low}$ &  $E_{\rm cut, high}^\dagger$ &    $E_{\rm fold, high}^\dagger$ &    $\Gamma_{\rm high}$ &         $A_{\rm high},10^{-3}$\\[2ex]

             &  $4.7_{-0.9}^{+1.3}$ &  $9.1_{-2.7}^{+3.0}$ &    $1.4_{-0.3}^{+0.2}$ &           &    $0.11_{-0.04}^{+0.05}$ &  $55.0_{-4.4}^{+8.3}$ &    $32.9_{-5.6}^{+6.3}$ &  $0.7_{-0.6}^{+0.5}$ &    $2_{-1}^{+6}$\\[1ex]\hline\\[-1.5ex]
\multirow{2}{*}{\cite{Coburn:2001p4571}} & $kT_{\rm bb}^\dagger$ & $E_{\rm cyc}^\dagger$ & $\sigma_{\rm cyc}^\dagger$ &  $\tau_{\rm cyc}$ &         $A_{\rm bb}$ &  $E_{\rm cut}^\dagger$ &    $E_{\rm fold}^\dagger$ &        $\Gamma$ &         $A_{\rm pl}$\\[1ex]
             & $1.69_{-0.07}^{+0.08}$ & $33.2_{-1.2}^{+1.2}$ &    $7.7_{-1.5}^{+9.9}$ & $5.5_{-1.5}^{+2.1}$ & $0.0027_{-0.0004}^{+0.0004}$ &  $67.5_{-5.8}^{+5.1}$ &   $48.7_{-9.5}^{+11.6}$ & $1.92_{-0.04}^{+0.04}$ &    $0.16_{-0.02}^{+0.03}$\\[1ex]\hline\\[-1.5ex]
\multirow{2}{*}{\textsl{CompTT + CompTT}} & $kT_{e {\rm , low}}^\dagger$ &       $\tau_{\rm low}$ &       $T_0^\dagger$ &           &         $A_{\rm low}$ & $kT_{e {\rm , high}}^\dagger$ &          $\tau_{\rm high}$ &     $T_0^\dagger$ &            $A_{\rm high}, 10^{-3}$\\[1ex]
             &  $4.6_{-0.8}^{+0.5}$ &  $9.2_{-0.7}^{+1.3}$ &   $0.83_{-0.06}^{+0.04}$ &           &  $0.036_{-0.005}^{+0.008}$ &  $15.3_{-0.5}^{+1.5}$ & $\ga 6$ & $0.83_{-0.06}^{+0.04}$ & $0.5_{-0.1}^{+0.7}$\\
	\end{tabular}
	\end{center}
	\caption{Best-fit results for the X-Per broadband spectrum obtained with \rm{INTEGRAL} and fitted with the discussed models. Here $kT_{bb,e,0}$ are the blackbody, electron or seed temperatures for the blackbody and Comptonization models, $\Gamma$ is the power-law photon index, $E_{\rm cut,fold}$ are the cutoff and fold energies for the cutoff power-law component. The ``low'' and ``high'' indices refer to the two spectral components. $^\dagger$-[keV]}
	\label{tab:allspe}
\end{table*}
\section{Discussion}
\paragraph{The broad band spectrum and the CRSF.}
First we briefly discuss the interpretation of the broadband spectrum and the
nature of the absorption feature proposed by \cite{Coburn:2001p4571}. As already stated, the double ``hump'' $Ef_E$ spectrum of the source
can be described by the models by
\cite{Coburn:2001p4571} and \cite{diSalvo:1998p4570}. We observe, however, that
the additional broad absorption-like feature suggested by
\cite{Coburn:2001p4571} is only required when a blackbody component is combined
with a relatively steep power law with a very hard cutoff. But this absorption line is not necessary when the low-energy part of the spectrum is
modeled with a power law rolling off at relative soft energies
\citep{diSalvo:1998p4570} or, in our preferred interpretation, with a
Comptonization model. The presence of the absorption feature is therefore model-dependent and its interpretation as a CRSF should be taken with caution. This
makes the estimate of the B field in the range of $10^{12}$\,G questionable.

\cite{diSalvo:1998p4570}, following early predictions of
\cite{Nelson:1995p5112}, suggested that the hard part of the spectrum is caused by
cyclotron \emph{emission} in the vicinity of the polar caps. In this scenario
the plasma is stopped on the neutron star polar caps via Couloumb collisions
at low accretion rates. A fraction (up to 5\%) of the kinetic energy of the
protons is transferred to the electron motions \emph{transverse} to proton
velocity (and magnetic field), thus leading to Landau level excitations. This
implies the formation of a broad ($E/\Delta E\sim2-4$) cyclotron
\emph{emission line}, which can be observed if photons are not thermalized in
the atmosphere of the neutron star. \cite{Coburn:2001p4571} criticized this
interpretation, arguing that the observed spectrum is not consistent with the
relatively narrow line predicted by \cite{Nelson:1995p5112}. In addition, we
observe that the cyclotron emission scenario is ruled out by energetic
considerations. From the observed broadband spectrum of X~Per we find that the
soft and hard components contribute $\sim$60\% and 30\% of the
total luminosity, respectively,\footnote{The hard component contributes only about
10\% of photons} with the hard component being a factor of 10 stronger
than that predicted by \cite{Nelson:1995p5112}.

To understand the nature of its hard component, it is interesting to compare
\xper with other accreting pulsars. At high accretion rates (above
$\sim10^{16}$\,g\,s$^{-1}$) the accretion flow in the vicinity of the neutron
star is stopped by the radiation pressure of the X-ray emission from the
pulsar in a so-called radiation-dominated shock with an accretion column
forming below this shock \citep{Basko}.
The X-rays emerging from
the polar caps are thermalized in the accretion column, producing the typical
cutoff power-law component observed in the spectra of accreting pulsars
\citep{Becker07}. At low accretion rates, as for \xper, the accretion column does
not form \citep{Basko}. However, the flow is optically thick along the
accretion direction. An accretion rate of $\sim 10^{15}$\,g\,s$^{-1}$ corresponds to
$\sim10^{29}$ particles\,cm$^{-2}$\,s$^{-1}$ for an accretion flow with radius
$R\sim500$\,m. On a time scale of $10^{-5}$\,s (or $\sim2$\,km assuming a
free-fall velocity of $\sim0.6c$) a column density of
$\sim10^{24}$\,particles\,cm$^{-2}$, equivalent to $\tau\sim1$ for Compton
scattering, can be reached. Soft photons emerging from the polar caps will
most likely scatter off relativistic electrons along the accretion flow and, therefore, one expects ``bulk'' Comptonization to play an important role in the 
spectral formation \citep{Becker05}. Note that at low accretion rates photons
can relatively easily escape perpendicularly to the accretion flow along the
side walls without being thermalized. \cite{Becker05, Becker07} were able to
qualitatively describe the observed spectrum of several sources including
\xper with purely bulk Comptonization, albeit in a limited energy range. Note
that a cutoff power law spectrum is still expected to emerge from the vicinity
of the polar cap, since thermal Comptonization is still expected to play a
role. This scenario can be described by the two spectral components we used
to fit the \rm{INTEGRAL} data.
This is of course a qualitative explanation and detailed spectral calculations
are needed. In particular, cyclotron emission should not be ignored, even if
it is not the dominant component. The pulse phase dependence of the spectrum
could certainly help in clarifying the suggested scenario. Unfortunately, the
statistics of the \rm{INTEGRAL} data did not allow us to constrain the phase-dependent parameters.
\paragraph{Link with other sources and the magnetic field} Dynamical or ``bulk''
Comptonization \citep{blandford81,psaltis06} has been invoked to describe
spectra of various sources, most notably the hard tails in the spectra of
low-mass X-ray binaries \citep{farin08} and accreting X-ray pulsars
\citep{Becker07}.
\cite{tor04} combined thermal and bulk Comptonization to describe the broadband
spectrum of \foru and predicted, based on their best-fit parameter, the
wind-accretion nature of the source. This prediction was later confirmed with
the discovery of pulsations with period $\sim5560$\,s \citep{reig09,finger10}.
\cite{finger10} determined that despite its long period and accretion rate, the
source spins down. This is possible only if the magnetic field is
$B\ge10^{15}$\,G, i.e., lies in magnetar range.
\xper and \foru are strikingly similar because they both accrete from wind, have a long spin
period and continue to spin-down on average. They have a similar spectrum and
luminosity, and as we have shown, there is likely no cyclotron line in \xper.
These consideration raise a key question: what is the magnetic field of \xper?
We can use the torque theory to get some insight on the anwer to this question. As we extensively discussed in \cite{megx}, following \cite{Illarionov:1990p1675}, the dipole component of the magnetic field of wind accreting X-ray pulsars can be estimated as
\begin{eqnarray}
B \approx 4\times10^{11}\,\mathrm{G}\left(\frac{k_w}{0.25}\right)^{7/8}\left(\frac{k}{2/3}\right)^{-7/8}\left(\frac{\xi}{0.87}\right)^{-7/8}\left(\frac{\dot{M}_\mathrm{eq}}{10^{15}\,\mathrm{g/s}}\right)^{1/2}\\ \nonumber
\left(\frac{\upsilon}{800\,\mathrm{km/s}}\right)^{-7/2} \left(\frac{P}{837\,\mathrm{s}}\right)^{7/8}\left(\frac{P_\mathrm{orb}}{250\,\mathrm{d}}\right)^{-7/8}\left(\frac{M}{1.4M_\odot}\right)^{2}\left(\frac{R}{10^6\,\mathrm{cm}}\right)^{-3},\nonumber
\label{eq:B}
\end{eqnarray}
here $k_w,k,\xi\simeq1$ are dimensionless coefficients, $\upsilon$ is the
velocity of stellar wind, $P,P_{\rm orb}$ are spin and orbital periods of
neutron star, and $M,R$ are mass and radius of the neutron star. For
\xper, fast accretion winds with a terminal velocity of $800$\,km\,s$^{-1}$
\citep{HammerschlagHensberge:1980p4826} would imply a magnetic field of
$B\sim10^{12}$\,G. On the other hand,
\cite{DelgadoMarti:2001p4574} argued that to explain the observed luminosity and
pulse period change rates of \xper, a slow wind subcomponent extending from the outskirts of the \emph{Be}, with velocity
$\upsilon\sim150$\,km\,s must be invoked.  A slow wind implies more captured orbital angular momentum
and therefore, from Eq.~1, a magnetic field in the ``magnetar'' range, i.e., $B\sim10^{14}$\,G. As discussed in \cite{megx}, other torque models found in
literature for wind accretion
\citep{Davidson:1973p2909,Davies:1979p2881,BisnovatyiKogan:1991p2029} give
similar values.

\cite{truemper10}, trying to link the quiescent emission of magnetars to that
of accreting pulsars, have suggested bulk Comptonization as the key mechanism
for the formation of the quiescent spectra of anomalous X-ray pulsars and
discussed the hard spectrum of \foru as the link between the two classes of
sources. Given the similarities with \foru, the same considerations can be
extended to \xper. That the torque theory for both accreting pulsars
suggests a magnetic field in the magnetar range makes this idea worthy of more
detailed investigation.

We note, however, that despite all similarities between the spectra of \foru,
\xper and AXPs (particularly 4U~0142+61), significant differences in the
spectra (see Fig.~\ref{fig:versus}), not discussed by \cite{truemper10}, are
evident. The spectrum of \axp extends to higher energies with no detectable
cutoff, while the soft component is much less prominent. A detailed
explanation of these differences is beyond the scope of this paper. However,
we believe that in the scenario outlined by \cite{truemper10}, the differences
can be qualitatively explained by the lower luminosity of \axp. Indeed, we
interpret the low-energy component as Comptonized emission from the vicinity
of the neutron star. Comptonization is more effective at higher densities,
meaning this component is likely more prominent for \xper, because it is
accreting at a significantly higher rate. On the other hand, the cutoff energy
of the high-energy component in the bulk Comptonization scenario depends on
the ratio between the average thermal and bulk energies \citep{farin08}. The
bulk energy of the flow is the same in both cases, while the lower luminosity
of \foru implies that less X-ray photons heat the plasma. Bulk Comptonization
is more important in this case and the cutoff shifts to higher energies.
Detailed calculations including the cyclotron cooling of the plasma are
required to confirm our qualitative scenario.

\section{Conclusions} We analyzed \rm{INTEGRAL} observations of the low-luminosity accreting pulsar X~Per. 
Our main conclusions are:
\begin{itemize}
	\item The source is significantly detected with \rm{ISGRI} above 100\,keV and shows the hardest spectrum among accreting pulsars.
	\item We successfully modeled, in line with previous findings, the 4-200\,keV broadband spectrum of the source with a two-component spectrum. No CRSF is necessary to model our data. We interpret the lower energy component as the result of thermal Comptonization. This component is, compared to other accreting pulsars, significantly reduced most likely because there is no accretion column. 
	\item The harder spectral component is interpreted for the first time as the result of the dynamical Comptonization in the accretion flow of photons emerging from the polar cap. 
	\item We showed that according to the current torque theory and taking into account the expected terminal wind velocity, a magnetic field of $\sim10^{14}$\,G has to be expected for \xper. 
	\item We discussed similarities of \xper with the slow accreting pulsar \foru and, as suggested, by \cite{truemper10} with \axp.
\end{itemize}

The low-luminosity, yet bright pulsars such as \xper and \foru can be good
laboratories to verify bulk Comptonization models because thermal
Comptonization is expected to be much less important than in high-luminosity
sources. Although we cannot conclude on the intenisty of the magnetic field of
the neutron star in X Per, we believe the ``high field'' scenario presented
above is coherent and worth further investigation. In this sense, \xper and
\foru are also ideal candidates for all efforts aiming at linking AXPs to
lower B-field accreting pulsars.
\begin{acknowledgements} V.D. thanks DFG for financial support
(grant DLR~50~OR~0702). \end{acknowledgements}
\vspace{-15pt}
\bibliography{biblio1} \end{document}